# Einstein's cosmology review of 1933: a new perspective on the Einstein-de Sitter model of the cosmos

Cormac O'Raifeartaigh,[a] Michael O'Keeffe,[a] Werner Nahm[b] and Simon Mitton[c]

[a]*School of Science, Waterford Institute of Technology, Cork Road, Waterford, Ireland*
[b]*School of Theoretical Physics, Dublin Institute for Advanced Studies, 10 Burlington Road, Dublin 2, Ireland*
[c]*Department of History and Philosophy of Science, University of Cambridge, Cambridge, UK*

Author for correspondence: coraifeartaigh@wit.ie

## Abstract

   We present a first English translation and analysis of a little-known review of relativistic cosmology written by Albert Einstein in late 1932. The article, which was published in 1933 in a book of Einstein papers translated into French, contains a substantial review of static and dynamic relativistic models of the cosmos, culminating in a discussion of the Einstein-de Sitter model. The article offers a valuable contemporaneous insight into Einstein's cosmology in the early 1930s and confirms that his interest lay in the development of the simplest model of the cosmos that could account for observation, rather than an exploration of all possible cosmic models. The article also confirms that Einstein did not believe that simplistic relativistic models could give an accurate description of the early universe.

## 1. Introduction

We recently came across a virtually unknown article written by Albert Einstein in late 1932 that contains a comprehensive review of static and dynamic relativistic models of the cosmos, culminating in a discussion of the Einstein-de Sitter model. The article, a signed, twelve-page handwritten manuscript titled *"Über das sogenannte kosmologische Problem"* (*"On the so-called cosmological problem"*), was found listed as document [1-115] on the Einstein Online Archive of the Hebrew University of Jerusalem (Einstein 1932a). It is our view that the document (figure 1) sheds useful light on Einstein's cosmology in the 1930s and on the Einstein-de Sitter model in particular.

According to the Albert Einstein Archive, manuscript [1-115] was sent by Einstein to his colleague Walther Mayer in early September 1932. Einstein collaborated closely with the young mathematician Mayer in the period 1930 to 1933 (Pais 1982 p492-493; Clark 1973 p 386, 391; Michelmore 1962 p161, 184) and they published many papers together on topics such as unified field theory (Einstein and Mayer 1930, 1931, 1932a), mathematics (Einstein and Mayer 1932b, 1934) and quantum mechanics (Einstein and Mayer 1933a, 1933b).[1] It is likely that the manuscript was sent to Mayer for review; indeed, one technical passage is in Mayer's handwriting, as discussed below.

Although Einstein's article was never published in a scientific journal, we have discovered that it was published in a little-known book of three Einstein papers translated into French by Einstein's lifelong friend and colleague Maurice Solovine. The article appeared under the title '*Sur la Structure Cosmologique de l'Espace*' in the book '*Les Fondaments de la Théorie de la Relativité Générale*'[2] published by Hermann in 1933 (figures 2 and 3). Correspondence between Einstein and Solovine in June 1932 indicates that the article was written specifically for the French book: *"I'm not sure which works you refer to…..but I could include a short treatise on my current approach to the cosmological problem"* (Einstein 1932b).[3] Einstein's manuscript was duly sent to Solovine on September 29th, 1932. An accompanying letter indicates that Einstein found his "short treatise" more substantial than expected, and intended to submit the work to a journal: *"You impatient scoundrel! I managed to tie the thing*

---

[1] The collaboration was so fruitful that, when negotiating his position at the Institute of Advanced Studies at Princeton, Einstein requested that a position also be found for Mayer (Pais 1982, p492-493; Isaacson 2007 p397).

[2] The book was one of a series of monographs by distinguished scientists published by Hermann et C$^{ie}$, an academic publishing house in Paris. The editorial board included Paul Langevin and Marie Curie, scientists admired by Einstein. The other two papers in the book were a translation of a classic paper on general relativity (Einstein 1916) and a translation of a paper on unified field theory (Einstein and Mayer 1931b).

[3] We thank Barbara Wolff of the Albert Einstein Archive for communicating this letter to us.

*together only after putting myself to a great deal of trouble and going through much reshuffling and some real work. But now it is crystal clear. I hope you will like it. However, I reserve the right to incorporate it later into an English publication that I have been promising for two years….Please return the manuscript after you have finished the translation"* (Einstein 1932c). In another letter a month later, Einstein proposed an alternative title for the article: *"I believe that we can change the title to "On the Structure of Space on the Largest Scales"* (Einstein 1932d).

We note that the promised "English publication" never appeared. It is likely that Einstein intended to publish the article in the *Monthly Notices of the Royal Astronomical Society*, given his friendship with Arthur Eddington (Vibert Douglas 1956, p100-102; Clark 1973 p398, Nussbaumer and Bieri 2009 p144) and given Eddington's role as papers secretary for the Royal Astronomical Society.[4] However, the paper did not appear in the *Monthly Notices,* or any other science journal, perhaps because the closing months of 1932 represented a time of great upheaval in Einstein's life.[5] Instead, the article was published only in a French booklet that did not enjoy a wide distribution,[6] and it was effectively lost to posterity.[7]

We present some historical remarks concerning Einstein's cosmology and the Einstein-de Sitter model in section 2 of this paper, followed by a guided tour of Einstein's article in section 3. As the document was published in 1933 (in French), we shall henceforth refer to it as Einstein's 1933 article (see Einstein 1933). We discuss the new insights offered by the article into the Einstein-de Sitter model in section 4, and conclude with some general remarks on Einstein's cosmology in these years in section 5. An English translation of the full text of Einstein's article is presented in an Appendix by kind permission of the Hebrew University of Jerusalem. Our translation is taken directly from Einstein's original handwritten manuscript in German (document [1-115] on the Einstein Online Archive); we have not found any points of disagreement with the French translation by Solovine. One page missing from Einstein's original manuscript is taken directly from the Solovine translation, as described in the Appendix.

---

[4] Eddington regularly sought submissions for the *Monthly Notices* from outstanding international figures. The most famous example was the republication of Lemaître's seminal 1927 paper in English (Lemaître 1931).
[5] With the victory of the National Socialists in the Reichstag in July 1932, Einstein's position in Germany became very uncertain (Clark 1973 p420; Michelmore 1962 p172-174; Pais 1994, p187; Isaacson, 396-399).
[6] Only a small number of copies of the book were issued and it soon went out of print.
[7] We are unaware of a single citation of the article during Einstein's lifetime and have found only two citations in historical reviews of cosmology (Kerzsberg 1989, p361-362; Eisenstaedt 1993, p106-107).

## 1. Historical context of the Einstein-de Sitter model

By the early 1930s, it had been established that the only static models of the cosmos allowed by general relativity (Einstein 1917; de Sitter 1917) presented some problems of a theoretical nature. De Sitter's empty universe was not truly static (Weyl 1923; Lemaître 1925), while Einstein's matter-filled universe was not stable (Lemaître 1927; Eddington 1930).[8] With Hubble's discovery of a linear relation between the recession of the galaxies and their radial distance (Hubble 1929), attention turned to the time-varying relativistic models of the cosmos that had been proposed independently by Alexander Friedman and Georges Lemaître in the 1920s (Friedman 1922; Lemaître 1927). A variety of cosmic models of the Friedman-Lemaître type were advanced in the early 1930s to describe Hubble's observations in terms of a relativistic expansion of space (Eddington 1930, 1931: de Sitter 1930a, 1930b; Tolman 1930a, 1930b, 1931, 1932; Heckmann 1931, 1932; Robertson 1932, 1933).[9] Einstein himself overcame his earlier distrust of time-varying models of the cosmos[10] and proposed two dynamic models during this period, the Friedman-Einstein model of 1931 and the Einstein-de Sitter model of 1932 (Einstein 1931a; Einstein and de Sitter 1932).

The Friedman-Einstein model (Einstein 1931a) marked the first paper in which Einstein formally abandoned his static model. Citing Hubble's observations, he took the view that his earlier assumption of a static universe was no longer justified: *"Now that it has become clear from Hubbel's* [sic] *results that the extra-galactic nebulae are uniformly distributed throughout space and are in dilatory motion (at least if their systematic redshifts are to be interpreted as Doppler effects), assumption (2) concerning the static nature of space has no longer any justification."* (Einstein 1931a).[11] Adopting Friedman's 1922 analysis of a universe of time-varying radius and positive spatial curvature,[12] Einstein also abandoned the cosmological constant he had introduced in 1917, on the grounds that it was now both unsatisfactory (it gave an unstable solution) and unnecessary: *"Under these circumstances, one must ask whether one can account for the facts without the introduction of the λ-term, which is in any case theoretically unsatisfactory"* (Einstein 1931a). The resulting model predicted a cosmos that would undergo an expansion followed by a contraction, and Einstein made use of Hubble's observations to extract estimates for the current radius of the universe, the mean density of matter and the timespan of the expansion.[13]

---

[8] See (Kerzberg 1989 p327-335) or (Smeenk 2015) for a review of the problems associated with the static models of Einstein and de Sitter.
[9] Most of these models assumed a positive curvature of space, following Friedman's analysis of 1922 and Lemaître's analysis of 1927.
[10] Einstein described Friedman's solution as *"hardly of physical significance"* in 1923, and dismissed Lemaître's model as *"abominable"* in 1927. See (Nussbaumer and Bieri 2009 p 92,111) or (Nussbaumer 2014a) for a review of Einstein's objection to dynamic models.
[11] We have recently given a first English translation of this paper in (O'Raifeartaigh and McCann 2014).
[12] Friedman's analysis included a cosmological constant (Friedman 1922).

In early 1932, Einstein and Willem de Sitter both spent time at Caltech in Pasadena, and they used the occasion to explore a new dynamic model of the cosmos (figure 4).This model took as starting point an observation by Otto Heckmann that the presence of a finite density of matter in a non-static universe did not necessarily imply a positive curvature of space - the curvature could also be negative or even zero (Heckmann 1931).[14] Mindful of a lack of empirical evidence for spatial curvature, Einstein and de Sitter set this parameter to zero: *"Dr O. Heckmann has pointed out that the non-static solutions of the field equations of the general theory of relativity with constant density do not necessarily imply a positive curvature of three-dimensional space, but that this curvature may also be negative or zero. There is no direct observational evidence for the curvature, the only directly observed data being the mean density and the expansion…and the question arises whether it is possible to represent the observed facts without introducing a curvature at all"* (Einstein and de Sitter 1932). With both the cosmological constant and spatial curvature removed, the resulting model described a cosmos of flat geometry in which the rate of expansion $h$ was related to the mean density of matter $\rho$ by the simple relation $h^2 = \frac{1}{3}\kappa\rho$, with $\kappa$ as the Einstein constant.[15] Applying Hubble's value of 500 km s$^{-1}$ Mpc$^{-1}$ for the recession rate of the galaxies to their model, the authors found that it predicted a value of 4x10$^{-28}$ g cm$^{-3}$ for the mean density of matter in the cosmos, a prediction they suggested was not incompatible with estimates from astronomy: *"Although, therefore, the density… corresponding to the assumption of zero curvature may perhaps be on the high side, it certainly is of the correct order of magnitude, and we must conclude that at the present time it is possible to represent the facts without assuming a curvature of three-dimensional space"* (Einstein and de Sitter 1932).

The Einstein-de Sitter model became very well-known and it played a significant role in the development of 20$^{th}$ century cosmology. One reason was that it marked an important hypothetical case in which the expansion of the universe was precisely balanced by a critical density of matter; a cosmos of lower mass density would be of hyperbolic spatial geometry and expand at an ever increasing rate, while a cosmos of higher mass density would be of spherical geometry and eventually collapse. Another reason was the model's great simplicity; in the absence of any observational evidence for spatial curvature or a cosmological constant, there was little reason to turn to more complicated models.[16] Indeed, the theory remained a favoured model of the universe for many years (North 1965 p134; Kragh 1996 p35; Nussbaumer and Bieri 2009 p 152; Nussbaumer 2014a), although it soon emerged that neither the predicted timespan of the expansion (see below) nor the required density of matter were in good agreement with observation.[17]

---

[13] We have recently suggested that these calculations contained some anomalies. We have also suggested that the model is not cyclic, as often stated (O'Raifeartaigh and McCann 2014).

[14] The possibility of negative curvature was explored by Alexander Friedman in 1924 (Friedman 1924) but this work was not widely cited for many years. Ironically, Friedman did not specifically consider the possibility of zero curvature in any of his works.

[15] Here the Einstein constant $\kappa$ was taken as $8\pi G/c^4$. The pressure of radiation was ~~also~~ assumed to be zero.

However, it is a curious fact that, while a detailed exposition of the Einstein-de Sitter *model* can be found in any modern textbook on cosmology, the *paper* published by Einstein and de Sitter in 1932 was an extremely terse affair. Noting that the only directly observable data were the mean density of matter and the expansion of the cosmos (see first quote above), the authors did not fully develop the model in the paper, but confined their interest to establishing a relation between these two parameters. Indeed, this approach is evident in the title of the paper – *"On the Relation between the Expansion and the Mean Density of the Universe"* (Einstein and de Sitter 1932). For example, the authors did not describe the time evolution of the expansion, an important aspect of any dynamic model of the cosmos. This omission was unfortunate, because the timespan of expansion implicit in the model was in fact problematic in comparison with estimates of the age of the earth from radioactivity, or estimates of the age of the stars from astrophysics (Kragh 1996 p 73-74: Nussbaumer and Bieri 2009 pp 153-155). The authors also passed over more general issues associated with the model, notably the puzzle of a matter-filled universe of infinite space,[18] and the problem of an implied beginning for spacetime.[19] Thus, it has often been noted that the Einstein-de Sitter paper of 1932 was a rather slight work (Kragh 1996 p35; Kragh 2007 p 156; Nussbaumer and

---

[16] Solid evidence for a positive cosmological constant did not emerge until 1992, while no evidence for spatial curvature has yet been detected.
[17] Astronomical observations measurements from 1932 onwards suggested a mean density of matter far below the critical value required by the Einstein-deSitter model (Oort J. 1932; Zwicky, F. 1933, 1937; Mitton 1976 pp 168-177).
[18] For example, Lemaître dismissed the possibility of a dynamic cosmos of Euclidean geometry in 1925 due to *"the impossibility of filling up an infinite space with matter which cannot but be finite"* (Lemaître 1925).
[19] The related question of an origin for the universe had been raised almost a year before (Lemaître 1931b,1931c).

Bieri 2009, pp150-152; Nussbaumer 2014a) [20] and it has even been suggested that the paper would hardly have been published had it been written by less illustrious authors (Nussbaumer and Bieri 2009, p150; Barrow 2011 p75).[21] We were therefore delighted to find a little-known review of relativistic cosmology written by Einstein later that same year, containing a detailed exposition of the Einstein de Sitter model.

## 2. A guided tour of Einstein's 1933 article[22]

Einstein begins his cosmology review by comparing the relativistic and pre-relativistic views of space and time. He points out that, in the non-relativistic case, space and time are seen as 'absolute' in the sense of a reference frame that has a tangible physical reality:

> *"When we call space and time in pre-relativistic physics "absolute", it has the following meaning. In the first instance, space and time, or the frame of reference, signify a reality in the same sense as, say, mass. Co-ordinates defined with respect to the chosen reference frame are immediately understood as results of measurement. Propositions of geometry and kinematics are therefore understood as relations between measurements that have the significance of physical assertions which can be true or false. The inertial reference frame is understood to be a reality because its choice is inherent in the law of inertia."*

Einstein then points out that space and time are viewed as 'absolute' in a second sense in non-relativistic physics, i.e., in the sense that they are not influenced the behaviour of material bodies:

> *"Secondly, in terms of the laws obeyed, the physical reality denoted by the words space and time is independent of the behaviour of the rest of the physically real world, that is, independent of material bodies for example. According to classical theory, all relationships between measurements, which can themselves only be obtained using rulers and clocks, are independent of the distribution and motion of matter; the same is true for the*

---

[20] Heckmann, who considered the case of zero curvature as one of many possible models in 1931 and 1932 (Heckmann 1931, 1932), described the Einstein-de Sitter paper was *"not very profound"* (Heckmann 1976, p 28).
[21] A well-known story by Eddington suggests that the authors themselves did not attach too much importance to the work at the time (Eddington 1940, p128; Plaskett 1933; Nussbaumer and Bieri 2009 p152).
[22] We suggest Einstein's article be read in full (Appendix) before reading this section.

> *inertial reference frame. Space enables the physical, in a sense, but cannot be influenced by the physical."*

Einstein opines that this view is not untenable, but it divides nature into an *a priori* reality and a secondary reality:

> *"Such a theory is by no means logically untenable, although it is unsatisfactory from an epistemological point of view. In it, space and time play the role of an* a priori *reality, as it were, different from the reality of material bodies (and fields) which appear to some extent as a secondary reality."*

He then points out that the general theory of relativity does not impose this division, citing this as a major achievement of relativity:

> *"It is precisely this unsatisfactory division of physical reality that the general theory of relativity avoids. From a systematic point of view, the avoidance of this division of physical reality into two types is the main achievement of the general theory of relativity."*

Einstein notes ~~that~~ another achievement of the general theory, namely that gravity and inertia can be described in a single framework:

> *"The latter also made it possible to comprehend gravity and inertia from a common perspective. In view of the above, the use of general Gaussian co-ordinates, which provide a continuous labelling of points in space-time without reference to metric relations, is a mere (albeit indispensable) tool that allows the metric properties of the continuum to be coordinated with its other properties (gravitational field, electromagnetic field, law of motion)."*

We note that in this interpretation, spacetime does not have an independent existence beyond giving expression to the relations among physical processes in the universe (see section 4). However, the new theory poses a puzzle not found in classical physics, namely the effect of a non-zero mean density of matter on the metric of spacetime, a puzzle Einstein names-the "so-called cosmological problem":

> *"Since, according to the general theory of relativity, the metric properties of space are not given in themselves but are instead determined by material objects that force a non-Euclidean character on the continuum, a problem arises that is absent from the classical theory. Namely, since we may assume that the stars are distributed with a finite density everywhere in the world, that is, a non-zero average density of matter in general, there arises the question of the influence of this mean density on the (metric) structure of* *space on a large scale; this is the so-called cosmological problem that we wish to address in this short note."*

To tackle the problem, Einstein first recalls the problem of gravitational collapse in a Newtonian universe, and a proposed solution by Hugo von Seeliger:

*"Moreover, the assumption of a finite non-zero mean density of matter already leads to difficulties from the point of view of the Newtonian theory, as the astronomers have long known. Namely, according to the theorem of Gauss, the number of lines of gravitational force that cross a closed surface from the outside to the inside is equal to a constant multiple of the gravitating mass enclosed by the surface. If this matter has the constant density ρ, then for a sphere of radius P, the number of lines of force is proportional to $P^3$. Therefore the flux of force per unit area of the sphere is proportional to the radius of the sphere, and so the greater the radius of the sphere, the greater it will be. Hence, according to Newton's theory, free matter of a finite constant density cannot remain in global equilibrium. To avoid the resulting difficulty, the astronomer Seeliger proposed a modification of the Newtonian law of attraction for large distances. Of course, this question had nothing to do with the problem of space."*[23]

Einstein then restates the problem for the case of general relativity:

*"The corresponding problem in the general theory of relativity leads to the question: how is it possible to have a space with a spatially constant density of matter that is at rest relative to it? Such a space shall be dealt with as the crudest idealization for a theoretical comprehension of the actual space-time-continuum."*

Einstein begins his analysis of the cosmological problem by recalling the field equations of general relativity:

*"According to the general theory of relativity, the metric or gravitational field described by the $g_{\mu\nu}$ is related to the energy or mass density tensor $T_{\mu\nu}$ by the equation*

$$R_{\mu\nu} - \frac{1}{2} g_{\mu\nu} R = -\kappa\, T_{\mu\nu} \quad \ldots (1)$$

---

[23] In order to avoid the problem of gravitational collapse in the Newtonian universe, Hugo von Seeliger suggested the introduction of an extra term to Newton's law of gravitation that would be effective only at the largest distances (Seeliger 1895, 1898). Einstein referred to Seeliger's solution in the third edition of his popular book on relativity (Einstein 1918 pp71-72) and in 1919, he stated that he would have cited Seeliger's solution in his cosmological model of 1917 had it been known to him at the time (Einstein 1919a).

*Here $R_{\mu\nu}$ signifies the once-contracted Riemann tensor*

$$R_{\mu\nu} = -\Gamma^{\alpha}_{\mu\nu,\alpha} + \Gamma^{\alpha}_{\mu\alpha,\nu} + \Gamma^{\alpha}_{\mu\beta}\Gamma^{\beta}_{\nu\alpha} - \Gamma^{\alpha}_{\mu\nu}\Gamma^{\beta}_{\alpha\beta} \quad ....1(a)$$

Making the assumption that influences such as the pressure of matter and radiation can be ignored, he constructs the stress-energy tensor in the usual manner:

*"If "matter" can be idealised as pressure-free, and the influence of effects other than gravity can be neglected, then*

$$T^{\mu\nu} = \rho u^{\mu} u^{\nu} \quad ... \quad (2)$$

*where $u^{\mu}$ denotes the contravariant four-dimensional velocity vector $\frac{dX_{\mu}}{d\tau}$ and $\rho$ the scalar density of matter. Naturally, it is also assumed that the energy density of the ponderable matter outweighs that of the radiation to the extent that the latter can be neglected. Although the validity of this assumption is not entirely assured, the approximation introduced does not essentially alter the results."*

Einstein first considers the simplest of all spacetime geometries, the static, flat spacetime of Minkowski, and notes that it is not compatible with a universe with a non-zero density of matter:

*"One sees first of all that a world with a non-zero density of matter cannot be euclidean. For such a world is given in terms of the special theory of relativity by a line element*

$$g_{\mu\nu} dx^{\mu} dx^{\nu} = ds^2 = dx_1^2 + dx_2^2 + dx_3^2 - c^2 dt^2 \quad ...(3)$$

*i.e., by constant values for the $g_{\mu\nu}$. $R_{\mu\nu}$ and $R$ then vanish and with them the left side of (1). It follows that the right-hand side of (1) must also vanish, and with it $\rho$, in contradiction to our assumption."*

Einstein then considers the next simplest metric, a static spacetime of constant curvature:

*"After euclidean space, the simplest spatial structure conceivable would seem to be one that is static (all $g_{\mu\nu}$ independent of t) and that has constant curvature with respect to the "spatial" sections (t = constant). As is well known, a three-dimensional space with constant positive curvature (in particular a "spherical" space) is characterised by the line element $d\sigma^2$ of the form:*

$$d\sigma^2 = \frac{dx_1^2 + dx_2^2 + dx_3^2}{\left(1 + \frac{r^2}{(2P)^2}\right)^2} \qquad (r^2 = x_1^2 + x_2^2 + x_3^2)$$

*where the point $x_1 = x_2 = x_3 = 0$ is only apparently singled out. A world that is static and spatially spherical is therefore described by the line element:*

$$ds^2 = \frac{dx_1^2 + dx_2^2 + dx_3^2}{\left(1 + \frac{r^2}{(2P)^2}\right)^2} - c^2 dt^2 \qquad \ldots (3a)"\ [24]$$

With the use of this metric, Einstein calculates values for $R_{\mu\nu} - \frac{1}{2}g_{\mu\nu}R$ and the stress-energy tensor $T_{\mu\nu}$, and derives two mutually contradictory differential equations:

> "*Thus, from (1) the two contradictory equations are obtained:*
>
> $$\frac{1}{P^2} = 0$$
>
> $$\frac{3c^2}{P^2} = \kappa\rho c^2 \qquad (4)$$
>
> *Therefore equations (1) do not allow the possibility of a non-zero uniform density of matter ρ. This immediately creates a serious difficulty for the general theory of relativity, given that time-independent spatial structures other than those given by (3a) (for positive or negative $P^2$) are inconceivable.*"

Einstein then recalls his 1917 solution to this problem, namely, the introduction of the cosmological constant to the field equations:

> "*I initially found the following way out of this difficulty. The requirements of relativity permit and suggest the addition of a term of the form $\lambda g_{\mu\nu}$ to the left hand side of (1), where λ denotes a universal constant (cosmological constant) which must be small enough that the additional term need not be considered in practice when calculating the sun's gravitational field and the motion of the planets. Completed in this manner, the equations are*
>
> $$\left(R_{\mu\nu} - \frac{1}{2}g_{\mu\nu}R\right) - \lambda g_{\mu\nu} = -\kappa T_{\mu\nu} \qquad \ldots (1b)$$
>
> *Instead of equations (4) one then finds:*

---

[24] We have obtained equation 3(a) and the sentence preceding it from the Solovine translation (Einstein 1933), as the passage containing this equation and the tensor calculations that follow is missing from manuscript [1-115] (see Appendix).

$$\left.\begin{array}{l}\frac{1}{P^2} = \lambda \\ \frac{3c^2}{P^2} = -\lambda c^2 + \kappa\rho c^2\end{array}\right\} \quad (4a)$$

*These equations are consistent and yield the following value for the world radius:*

$$P = \frac{2}{\sqrt{\kappa\rho}} = \frac{c}{\sqrt{2\pi K\rho}} \qquad (5)$$

*where* K *denotes the gravitational constant as measured in the usual system of measurement."*

Einstein then points out that this solution was subsequently found to be unstable, citing the work of Friedman and Lemaître:[25]

*"However, it later emerged as a result of research by Lemaitre and Friedmann that this resolution of the difficulty is unsatisfactory for the following reason. The above-mentioned authors… generalized the approach (3a) by introducing the world radius P (and the density ρ) not as a constant, but rather as an* a priori *unknown function of time. Equations (1b) then show that solution (4a),(5) has an unstable character."*

Addressing the new time-dependent models, Einstein first notes that the magnitude and sign of the cosmological constant are no longer determined, nor is the sign of spatial curvature:

*"Furthermore, if one adopts these "dynamic" solutions… then both the magnitude and the sign of λ will remain undetermined, and indeed even the sign of $\frac{1}{P^2}$, so that negative spatial curvatures also appear possible and thus the basis for the postulate of a spatially closed world is completely removed."*

At this point, Einstein cites Otto Heckmann as the first to suggest the possibility of negative spatial curvature (see appendix).[26] We note that Einstein is still apparently unaware that Friedman explored the possibility of negative spatial curvature many years before (Friedman 1924). Before addressing the issue of curvature further, Einstein suggests that the dynamic models have rendered the cosmological constant redundant:

> *"Given that the theory leads us to adopt dynamic solutions for the structure of space, it is no longer necessary to introduce the universal constant λ, as there are dynamic solutions for (1) of type (3a) for which $\lambda = 0$."*

---

[25] This is the first time that Einstein cites Lemaître's dynamic model in a scientific paper.
[26] Einstein does not give a specific reference for Heckmann's work. It is likely that he is referring to Heckmann's cosmology paper of 1931 (Heckmann 1931), rather than Heckmann's more comprehensive paper of 1932 (Heckmann 1932) as Einstein's review was written in late 1932.

In the next paragraph, Einstein turns to observation, providing a succinct summary of the discovery by astronomers of a velocity/distance relation for the spiral nebulae. He notes that the astronomical observations gave a new impetus to dynamic cosmologies, and offered support for the assumption of cosmic homogeneity:

> *"In recent times, the resolution of the problem has received a strong stimulus from empirical results in astronomy. Measurements of the Doppler effect (in particular those of Hubbel) of the extra-galactic nebulae, which have been recognized as similar formations to the Milky Way, have shown that the further these formations are from us, the greater the velocity with which they hasten away. Hubbel's investigations also showed that these formations are distributed in space in a statistically uniform manner, giving empirical support to the underlying theoretical assumption of a uniform mean density of matter. The discovery of the expansion of the extra-galactic nebulae justifies the shift to dynamic solutions for the structure of space, a step that heretofore would have appeared to be an expedient justified only by theoretical necessity."[27]*

We note that Einstein cites only Hubble, although the pioneering redshift observations of Vesto Slipher played a critical role in the discovery of a redshift/distance relation for the nebulae.[28] This omission is a consistent feature of Einstein's writings on dynamic cosmology (Einstein 1931a, Einstein and de Sitter 1932, Einstein 1945) and may have been a factor in the overlooking of Slipher's contribution to the discovery of the expanding universe (O'Raifeartaigh and McCann 2014).

Returning to the theoretical models, Einstein now considers the case of time-dependent models of constant spatial curvature, and sets the cosmological constant to zero. He notes that such models can be neatly described in terms of an expanding scale factor $\frac{P}{P_0}$:

---

[27] The German text reads "*die Expansion der extragalaktischen Nebel*", but Einstein presumably meant the recession of the nebulae.
[28] It has been shown that over two thirds of the redshift data used by Hubble in 1929 were from Slipher (O'Raifeartaigh 2013; Peacock 2013; Way 2013).

> *"Thus, the theory can now, without the introduction of the λ term, accommodate a finite (mean) density of matter ρ on the basis of equations (1) by means of the relation 3(a) with P (and ρ) variable over time. Here, it should be noted that it is not the coordinates $x_1, x_2, x_3$ of a particle that remain constant over time, but the quantities $\frac{x_1}{P}, \frac{x_2}{P}, \frac{x_3}{P}$, as is seen from a straightforward geometric argument. We will not introduce these quantities themselves as new coordinates, but instead the quantities $P_0 \frac{x_1}{P}, P_0 \frac{x_2}{P}, P_0 \frac{x_3}{P}$, where $P_0$ denotes a length of the order of magnitude of the "world radius". We do this in order to ensure that differences between coordinates will be of the same order of magnitude as lengths measured with a ruler. If we again label each of these new co-ordinates $x_1, x_2, x_3$ and in this new system again let $r = \sqrt{x_1^2 + x_2^2 + x_3^2}$, then the relation (3a) takes the form:*
>
> $$ds^2 = \left(\frac{P}{P_0}\right)^2 \frac{dx_1^2 + dx_2^2 + dx_3^2}{\left(1 + \frac{r^2}{(2P_0)^2}\right)^2} - c^2 dt^2 \quad \ldots (3b)$$
>
> *We can regard $P_0$ is the world radius P at a particular point in time $t_0$. Only the "expansion factor" $\frac{P}{P_0}$ ( = A) is then variable over time."*

We note that this model is somewhat similar to Einstein's cosmic model of 1931 (Einstein 1931a). However, the treatment here is more general, as Einstein derives the line element for the case of constant curvature from first principles and does not specify whether the curvature is positive or negative.

In the last section of the manuscript, Einstein notes that the presence of a finite density of matter in a dynamic cosmos does not automatically imply a curvature of space:

> *"We have already noted that, if we take A to be constant over time, i.e., without an expansion of space, we cannot explain a constant density of matter ρ solely by the assumption of a curvature of space. On the other hand, it will be shown that the existence of a finite density ρ does not in any way demand the existence of a (3-dimensional) curvature of space."* [29]

He thus proceeds to analyse the case of a dynamic cosmic model with both spatial curvature and the cosmological constant set to zero, i.e., $k=0$, $\lambda=0$. With the use of this metric, Einstein derives two differential equations (see figure 2) from the field equations:

---

[29] Heckmann is cited as the first to consider this possibility in (Einstein and de Sitter 1932), but the attribution is not repeated here.

> *"This amounts to replacing (3b) by:*

$$ds^2 = A^2(dx_1^2 + dx_2^2 + dx_3^2) - c^2 dt^2 \quad ... \quad (3c)$$

*where A is a function of t (= $x_4$) alone. Introducing this relation in (1) gives*

$$2A \frac{d^2A}{dt^2} + \left(\frac{dA}{dt}\right)^2 = 0 \quad ... \quad (6)$$

$$3\left(\frac{\frac{dA}{dt}}{A}\right)^2 = \kappa\rho c^2 \quad ... \quad (7)\text{"}$$

Thus, Einstein has derived two differential equations, analogous to the Friedman equations, for the special case of a cosmos of flat geometry and vanishing cosmological constant. We note that equation (7) is almost identical to that of the Einstein-de Sitter paper,[30] while equation (6) was omitted in that article (Einstein and de Sitter 1932).

From equation (6), Einstein develops an expression for the timespan of the expansion, noting that equation (7) implies an infinite density of matter at some point in the past:

> *"Equation (6) yields:*
>
> $$A = (t - t_0)^{2/3} \quad (6a)$$
>
> *If l is the time-independent distance $\sqrt{\Delta x_1^2 + \Delta x_2^2 + \Delta x_3^2}$ between two masses... then according to (3c), Al is the distance D between these two mass points as measured with a ruler. (6a) thus expresses an expansion that begins at a particular point in time $t_0$. For this point in time (7) shows that the density is infinite."*

Einstein then notes that Hubble's observations give a value for the expansion factor, and uses it to estimate the timespan of the expansion. He does not state explicitly what value he assumes for the Hubble constant, but the calculation of matter density that follows implies the same value used in the Einstein-de Sitter model (500 kms$^{-1}$ Mpc$^{-1}$).[31]

---

[30] There is a difference of a factor of $c^2$ between the two expressions, arising from a difference in units for the Einstein constant $\kappa$.
[31] The standard value at the time was 500 km s$^{-1}$ Mpc$^{-1}$ (Hubble 1929; Lemaître 1931a,1933; Eddington 1931a; Einstein and de Sitter 1932; Kragh 2007 p160).

> *"Hubbel's measurements of the extra-galactic nebulae have shown that for the present, $\frac{1}{D}\frac{dD}{dt}(=\frac{1}{A}\frac{dA}{dt})$ is a constant h. If t is the present time, then according to (6a)*
>
> $$t - t_0 = \frac{2}{3h} \quad \ldots (8)$$
>
> *This time-span works out at approximately $10^{10}$ years."*

We note that this approximation is somewhat inaccurate; in fact equation (8) implies a time of 1.3 billion years for the expansion, as discussed in section 4. Einstein's figure is nonetheless lower than the ages of the stars estimated from astrophysics at the time (Condon 1925, Jeans 1928, p381). He does not comment specifically on the paradox, but points out that the model can be expected to fail at early epochs because the assumption of homogeneity is likely to be invalid:

> *"Of course, at that time, the density will not actually have been infinitely large: Laue has rightly pointed out that our rough approximation, according to which the density ρ is independent of location, breaks down for this time."*[32]

Finally, Einstein turns his attention to equation (7), a relation between the cosmic expansion and the mean density of matter. He uses the Hubble constant to estimate a value for the matter density in the same manner as the published Einstein-de Sitter paper (Einstein and de Sitter 1932), noting once again that the estimate is not incompatible with astronomical observation:

> *"Applying (7) to the present yields*
> $$3h^2 = \kappa\rho c^2 \ (= 8\pi K\rho)\ldots\ldots\ldots(9)$$
> *This is a relation between the Hubbel constant h, determined from the Doppler effect, and the mean density ρ. Numerically, this equation gives an order of magnitude of $10^{-28}$ for ρ, which is not incompatible with the estimates of the astronomers."*

Units of measurement are not stated for the density ρ, but simple calculation or comparison with the published Einstein-de Sitter paper shows that the units are g cm$^{-3}$.

---

[32] A close colleague of Einstein's during the Berlin years, Max von Laue made many important contributions to physics, including an early textbook on general relativity and gravitation (Laue 1921). Einstein is probably referring to a 1931 article by Laue in which he noted that the assumption of a homogenous distribution of matter was not justified by observation (Laue 1931).

Einstein concludes his article by restating the main result of his analysis, namely that the presence of a finite density of matter in an expanding universe does not necessarily demand a curvature of space. He is careful to point out that this result does not in itself imply that we inhabit a universe of flat geometry (although there is at present no evidence of spatial curvature) and concludes by suggesting that spatial curvature might exist on a smaller scale than he envisioned in 1917:

> *"It follows from these considerations that in the light of our present knowledge, the fact of a non-zero density of matter need not be reconciled with a curvature of space, but instead with an expansion of space. Of course, this does not mean that such a curvature (positive or negative) does not exist. However, there is at present no indication of its existence. In any case, it may well be substantially smaller than might have been suggested by the original theory (see equation 5)".*

## 4. Discussion

It is evident from section 3 above that Einstein's 1933 article offers a much more substantive discussion of relativistic cosmology than his other papers on cosmology in these years (Einstein 1931a, Einstein and de Sitter 1932). The document thus provides a valuable contemporaneous insight into Einstein's cosmology at a critical moment in 20[th] science, the discovery of the first empirical evidence for an expanding universe.

We note first Einstein's philosophical discussion of the relativistic and non-relativistic views of space and time in the opening paragraphs of the article, the only such discussion in Einstein's papers on cosmology. Of particular interest is his observation that, in Newtonian physics, space and time have a reality that is different from the reality of ordinary objects, a schism that is removed by relativity: *"In it, space and time play the role of an* a priori *reality, as it were, different from the reality of material bodies (and fields) which appear to some extent as a secondary reality. It is precisely this unsatisfactory division of physical reality that the general theory of relativity avoids."* A similar discussion can be found in other accounts of general relativity by Einstein (Einstein 1918 pp 94-95; Einstein 1949 pp 65-66). Further on, Einstein notes that space and time do not have an existence independent of matter in the general theory: *"The metric properties of space are not given in themselves but are instead determined by material objects that force a non-Euclidean character on the continuum".* This 'relational' interpretation of spacetime is a consistent feature of Einstein's writings on general relativity and can be found in many of his reviews of the subject (Einstein 1918 p95; Einstein 1945 pp57-58; Einstein 1949 p65).

Einstein continues his philosophical discussion by considering the problem of gravitational collapse in Newtonian physics and the corresponding 'cosmological problem' in general relativity: *"The assumption of a finite non-zero mean density of matter already leads to difficulties from the point of view of the Newtonian theory... the corresponding problem in the general theory of relativity leads to the question: how is it possible to have a space with a spatially constant density of matter that is at rest relative to it?"* This interesting comparison is absent in the Einstein-de Sitter paper (Einstein and de Sitter 1932), although a similar discussion can be found in the Friedman-Einstein model of 1931 (Einstein 1931a).

To address the cosmological problem, Einstein then reviews five models of the cosmos: (i) a stationary cosmos of flat geometry (ii) a stationary cosmos of constant spatial curvature (iii) a stationary cosmos of constant spatial curvature with a cosmological constant (iv) a dynamic cosmos of constant spatial curvature without a cosmological constant (v) a dynamic cosmos without spatial curvature or cosmological constant. It is demonstrated from first principles that models (i) and (ii) are not compatible with a non-zero density of matter, while model (iii) leads to an unstable solution. Model (iv) comprises a generalization of Einstein's model of 1931 (Einstein 1931a) to include negative spatial curvatures. Finally, Einstein considers the case of a matter-filled, expanding cosmos of Euclidean geometry (model v). This model was proposed by Einstein and de Sitter in April 1932, but it is more fully developed in Einstein's 1933 review, in a manner that sheds useful light on his view of issues such as the cosmological constant, the curvature of space and the timespan of dynamic models of the universe.

### *On the cosmological constant*

It is a staple of many accounts of 20[th] century cosmology that Einstein introduced the cosmological constant to the field equations in 1917 in order to predict a static rather than a dynamic universe.–However, it is probably more accurate to say that the purpose of the cosmological constant was to allow the prediction of a non-zero density of matter in a universe that was assumed *a priori* to be static (Einstein 1917). Indeed, the notion of a time-varying universe would have seemed very far-fetched at the time (North 1965 p82; Kragh 1996 p9; Kragh 2007 p134). Assuming a cosmos that was static over time, and that a consistent theory of gravitation should incorporate Mach's principle, Einstein noted that a new term comprising the fundamental tensor $g_{\mu\nu}$ multiplied by a universal constant λ could be added to the field equations without destroying the general covariance, giving a static universe of finite matter density and closed curvature (neatly avoiding the problem of boundary conditions).[33] This subtle point is made abundantly clear in the 1933 article. Comparing equations (4) with (4a), Einstein remarks: *"Therefore, equations (1) do not allow the possibility of a non-zero uniform density of matter ρ. This immediately creates a serious difficulty for the general theory of relativity, given that time-independent spatial structures other than those given by (3a) (for positive or negative $P^2$) are inconceivable."*

A great deal has been written over the years about Einstein's view of the cosmological constant. We will not review that literature here,[34] but note that he clearly disliked the term from the point of view of relativity, commenting already in 1919 that it was *"gravely detrimental to the formal beauty of the theory"* (Einstein 1919b). We also note that Einstein proposed a reformulation of the field equations that year in which the cosmological constant appeared as a constant of integration, rather than a universal constant peculiar to the law of gravitation: *"..the new formulation has this great advantage, that the quantity appears in the fundamental equations as a constant of integration, and no longer as a universal constant peculiar to the fundamental law"* (Einstein 1919b).

With the emergence of the first evidence for an expanding universe, Einstein removed the cosmological constant term on the grounds that it was now both unsatisfactory (it gave an unstable solution)[35] and unnecessary. As he commented in his cosmic model of 1931: *"Under these circumstances, one must ask whether one can account for the facts without the introduction of the λ-term, which is in any case theoretically unsatisfactory"* (Einstein 1931a).[36] In his 1933 article, Einstein expands on this point. First, the cosmological constant is banished on theoretical grounds: *"Equations (1b) then show that solution (4a),(5) has an unstable character….given that the theory leads us to adopt dynamic solutions for the structure of space, it is no longer necessary to introduce the universal constant λ, as there are dynamic solutions for (1) of type (3a) for which λ = 0."* Second, the term has been rendered redundant by observation: *"The discovery of the expansion of the extra-galactic nebulae justifies the shift to dynamic solutions for the structure of space, a step that*

---

[33] Einstein's view of Mach's principle in these years was that space could not have an existence independent of matter; thus the spatial components of the metric tensor should vanish at infinity (Einstein 1917; Earman 2001)

[34] See (Earman 2001), (Straumann 2002) or (Smeenk 2014) for a recent review.

[35] It was shown in 1930 that the solution was unstable against the slightest perturbation in the density of matter (Eddington 1930).

[36] An intriguing anticipation of this action can be seen in Einstein's response to Weyl's analysis of the de Sitter model (Weyl 1923):*"..if there is no quasi-static world, then away with the cosmological term"* (Einstein 1923: Pais 1982 p288).

*heretofore would have appeared to be an expedient justified only by theoretical necessity. Thus the theory can now, without the introduction of a λ-term, accommodate a finite (mean) density of matter ρ on the basis of equations (1), using relation 3(a) with P (and ρ) variable over time."* Thus Einstein took the view that a term that was no longer necessitated by theory or observation had no place in relativistic cosmology. This view was not shared by many of his colleagues. One reason was that the inclusion of the cosmological constant in the field equations constituted the most general form of the theory; and while models with a non-zero $\lambda$ might not be necessary to account for observation, they could not be ruled out on the basis of empirical evidence (Lemaître 1927, 1931a; Eddington 1930; Robertson 1932, 1933; Heckmann 1931, 1932; de Sitter 1932 p126-127; Tolman 1931b, 1934 p482). Another reason was that the term could play a role in addressing the problematic timespan of expanding models (Eddington 1930; Lemaître 1931c, 1934; de Sitter 1933). Still other reasons were that the term could give a physical cause for cosmic expansion (Eddington 1931a, 1931c, 1933 p23-24; de Sitter 1931; Lemaître 1934) or could play a role in the formation of structure in the expanding universe (Lemaître 1933). Einstein was not swayed by any of these arguments and did not subsequently change his view on the matter. His attitude to the cosmic constant is probably best summarized in his 1945 article on cosmology: *"If Hubble's expansion had been discovered at the time of the creation of the general theory of relativity, the cosmologic member would never have been introduced. It seems now so much less justified to introduce such a member into the field equations, since its introduction loses its sole original justification – that of leading to a natural solution of the cosmologic problem."* (Einstein 1945 p130).[37]

### *On the curvature of space*

The opening paragraph of the Einstein-de Sitter paper contains the statement *" Non-static solutions of the field equations of the general theory of relativity with constant density do not necessarily imply a positive curvature of three-dimensional space…"* and the authors ask *" whether it is possible to represent the observed facts without introducing a curvature at all"* (Einstein and de Sitter 1932). As discussed in section 2, it is sometimes argued that the paper was a rather slight work. However, the concluding statement of Einstein's 1933 article emphasizes the true purpose of the Einstein-de Sitter model, namely the clarification of an important theoretical point in relativistic cosmology: *"It follows from these considerations that the fact of a non-zero density of matter need not be reconciled with a curvature of space, but instead with an expansion of space"*. The point is not that Einstein is convinced that we inhabit a universe of Euclidean geometry, but that one must guard against the notion that a curvature of space is implied by a finite density of matter – this is not necessarily the case for a dynamic universe.[38] More generally, it is often forgotten that most of the cosmic models proposed in the years 1930-1932 assumed a closed spatial geometry (Eddington 1930, 1931a; de Sitter 1930; Tolman 1930a, 1930b, 1931, 1932), as they were based on the earlier analyses of Friedman and Lemaître (Friedman 1922: Lemaître 1927).[39] The Einstein-de Sitter model was an important step forward in theoretical cosmology because it represented the first well-known open, infinite model. With that in mind, it is a little disappointing that Einstein does not discuss the apparent paradox of a matter-filled universe of infinite space in his 1933 article, given his long interest in space, time and matter.[40]

---

[37]One exception to this narrative is that Einstein retained the cosmological constant in an attempt at a steady-state model of the cosmos; however, he abandoned the model before publication (Einstein 1931b; O'Raifeartaigh et al 2014; Nussbaumer 2014b).

A second motivation for the Einstein-de Sitter model can be discerned from the 1933 review, namely the avoidance of unnecessary complications in relativistic cosmology. The document makes clear Einstein's thought progression from a static to a dynamic universe; from a dynamic universe of constant curvature and cosmological constant to a dynamic universe without $\lambda$; and finally to a dynamic universe without $\lambda$ and without curvature. This approach indicates that, far from taking a perfunctory interest in cosmology, Einstein is engaged in a systematic search for the simplest possible model of the cosmos that can account for observation. If neither the cosmological constant nor spatial curvature are necessary in order to describe the observed universe, why include them? This 'Occam's razor' approach is a consistent feature of Einstein's cosmology. His interest is not in the development of a general framework for all possible models of the universe, in the manner of Friedman, Heckmann, Robertson or Tolman (Friedman 1922, 1924; Heckmann 1931, 1932; Robertson 1933; Tolman 1934 p394, 403), but rather in the simplest model that can account for observation. This approach is very reminiscent of the young Einstein's pragmatic approach to emerging phenomena in physics (Einstein 1905a, 1905b, 1905c).

### *On the timespan of the expansion*

---

[38] However, Einstein does not dismiss the possibility of spatial curvature, unlike the case of the cosmological constant.
[39] Friedman's exploration of negative spatial curvature (Friedman 1924) was not widely cited at the time.
[40] See for example (Einstein 1918 pp71-72) or (Friedman 2014) for a review.

A significant difference between Einstein's 1933 review and the Einstein-de Sitter paper is that the time dependence of the model is fully analyzed in the review, while this topic was omitted in the Einstein-deSitter paper (Einstein and de Sitter 1932). (As pointed out in section 2, the latter omission was unfortunate because the timespan implied by the Einstein-de Sitter model was problematic in comparison with estimates of the age of stars from astrophysics). We noted in section 3 that in his 1933 article, Einstein derives the two differential equations (6) and (7) from the field equations, and extracts from (6) the relation $A = (t - t_0)^{2/3}$ between the coefficient of expansion $A$ and the timespan of the expansion $t - t_0$. With the use of the Hubble constant,[41] he then determines a value of 10 billion years for the timespan. In fact, this value should be 1.3 billion years, as pointed out in section 3. We have recently noted a similar inaccuracy in Einstein's model of 1931 (O'Raifeartaigh and McCann 2014). One explanation may be that Einstein is not familiar with calculations involving the astronomical units of megaparsecs and is simply repeating Friedman's generic estimate of the timespan of dynamic models of the cosmos (Friedman 1922). Another explanation may be that Einstein sees the figure of 10 billion years as a rough upper bound, including uncertainties in the determination of the Hubble constant. However, an overestimate by an order to magnitude seems somewhat extravagant. Given that Einstein calculates a timespan of 1.5 billion years for the Einstein-de Sitter model a decade later (Einstein 1945 p 124), it seems likely that a numerical error has occurred.[42]

The relatively short timescale of expanding models (with or without spatial curvature) was widely recognized as a serious difficulty for relativistic cosmology in these years (Eddington 1930; Tolman 1934 pp 485-486; de Sitter 1933; Lemaître 1934: North 1965 pp 223-229; Kragh 1996 pp 73-79). One solution was to augment the age of the universe using the cosmological constant (Eddington 1930; Lemaître 1931c, 1934; de Sitter 1933) and one wonders how the problem could be addressed in a model without the term. Einstein's 1933 article provides an answer to this question; as seen in section 3, he assumes the model will not be reliable at early times because it is unlikely that the simplifying assumption of a homogeneous distribution of matter will be justified: *"Laue has rightly pointed out that our rough approximation, according to which the density ρ is independent of location, breaks down for this time."* We note that Einstein employed the same argument more explicitly in the case of his model of 1931 : *"The greatest difficulty with the whole approach, .. is that… the elapsed time since P = 0 comes out at only about $10^{10}$ years. One can seek to escape this difficulty by noting that the inhomogeneity of the distribution of stellar material makes our approximate treatment illusory."* (Einstein 1931a). With Einstein's more accurate calculation of the timespan of expanding models in 1945, a more general caution is added: *"For large densities of field and of matter, the field equations and even the field variables which enter into them will have no real significance. One may not therefore assume the validity of the equations for very high density of field and of matter"* (Einstein 1945 p132-133). Thus, it is clear that Einstein attributed the problematic timespan of relativistic models of the cosmos to the simplifying assumptions made in the models. Ironically, it was later discovered that the problem lay in astronomical observation.[43]

---

[41] Einstein's calculation of the density of matter implies that he assumed a value of 500 kms$^{-1}$ Mpc$^{-1}$ for the Hubble constant, the standard value at the time.

[42] Indeed, Einstein notes that the estimate is *"a paradoxical result"* when compared with estimates of the age of the earth from radioactivity (Einstein 1945 p124).

### *On the origin of the universe: the dog that didn't bark*

To modern eyes, a striking aspect of Einstein's 1933 review, his most substantial discussion of dynamic cosmology in these years, is the lack of a discussion of the problem of the singularity, or of the related question of an origin for the universe. This omission seems curious, given Lemaître's hypothesis of a 'fireworks beginning' for the universe over a year before (Lemaître 1931b, 1931c). However, Einstein's silence on the issue is very typical of his cosmology in these years – there is no reference to the question of cosmic origins in the Friedman-Einstein model (Einstein 1931a), in the Einstein-de Sitter paper (Einstein and de Sitter 1932), or even in an unpublished exploration of a steady-state model of the expanding cosmos (Einstein 1931b).[44] Einstein's 1933 review offers a simple explanation for this silence; as noted above, he had little confidence in the accuracy of simplistic relativistic models of the cosmos extrapolated to early epochs.[45]

One should not perhaps not conclude that Einstein was necessarily opposed to the idea of a beginning of the world in the sense of an origin for the stars and the galaxies. In 1933, when Lemaître proposed at a seminar at Caltech that cosmic rays could represent the remnants of a 'fireworks beginning' of the universe, Einstein is reported to have lauded the idea as the *"most pleasant, beautiful, and satisfying interpretation of the source of cosmic rays that has been presented"* (AP 1933: Aikman 1933; Kragh 1999 p408; Farrell: 2005 pp101-102). It may thus be the case that Einstein was not adverse to the notion of a beginning for the physical world *per se*; he was simply distrustful of assertions based on the extrapolation of simplified relativistic models to early epochs.[46] This attitude can also be found in Einstein's later writings on cosmology: *"One may not therefore assume the validity of the equations for very high density of field and of matter…This consideration does not however alter the fact that the 'beginning of the world' really constitutes a beginning, from the point of view of the development of the now existing stars and systems of stars"* (Einstein 1945 p133).

---

[43] In the 1950s, it was discovered that Hubble had significantly underestimated the distances to the galaxies. New observationsby Baade and Sandage (Baade 1952; Sandage 1958) gave a much reduced value for the Hubble constant. By 1958, the timescale of the Einstein-de Sitter model was estimated to be 7-13 billion years (Kragh 1996 p 272-273).

[44] We have recently given a translation and analysis of this unpublished work (O'Raifeartaigh et al 2014). While the problematic timespan of evolving models is cited as a motivation for a steady-state solution, the problem of origins is not mentioned.

[45] This reluctance to speculate on the question of cosmic origins was by no means untypical in these years (Eddington 1933 p124-126; de Sitter 1932 pp131-133; Tolman 1934 p484-486; Robertson 1932, 1933).

## 5. *Concluding remarks*

Einstein's 1933 article provides a valuable contemporaneous insight into his cosmology in the 1930s. The work represents Einstein's only review of cosmology in these years, and supports our previous understanding of his views on the cosmological constant and on the curvature of space. The article confirms that Einstein's interest lay in the development of the simplest model of the cosmos that could account for observation, rather than an exploration of all possible cosmic models. The article also confirms that Einstein did not believe that simplified relativistic models could give an accurate description of the early universe or address the question of origins.

It is likely that Einstein's original title for the paper, *'On the so-called cosmological problem'* (Einstein 1932a) was chosen to convey a view that the cosmological problem was now solved, i.e., that the question of the influence of a finite density of matter on the structure of space no longer presented a theoretical conundrum now that it was accepted that time-varying geometries were possible.[47] Some support for this conclusion can be found in Einstein's letter to Solovine in October 1932: *"I inserted the word 'so-called' into the expression 'Cosmological Problem' because the title did not accurately characterize the subject dealt with. I believe that we can change the title to "On the Structure of Space on the Largest Scales"* (Einstein 1932d).[48] This observation may also explain why Einstein did not publish any cosmic models after 1933. As the structure of space in relativistic models no longer presented a theoretical puzzle, there was little point in exploring further models of the cosmos in the absence of empirical values for key cosmological parameters such as spatial curvature and the density of matter. As Einstein commented in his 1945 essay on cosmology: *"It seems we have to take the idea of an expanding universe seriously, in spite of the short 'lifetime'…. If one does so, the main question becomes whether space has positive or negative spatial curvature…an empirical decision does not seem impossible at the present state of astronomy. Since h (Hubble's expansion) is comparatively well known, everything depends on determining ρ with the highest possible accuracy"* (Einstein 1945 p133).

---

[46] We note that Einstein suggested to Lemaître in 1933 that effects such as anisotropy and inhomogeneity should be incorporated into relativistic models of the cosmos (Lemaître 1933; Lemaître 1958).

[47] Einstein also used the expression 'so-called cosmological problem' in his later review of cosmology (Einstein 1945, p113).

[48] We note that Einstein's suggested title *"Über die Struktur des Räumes im Grossen"* is mistranslated in (Einstein 1932d) as *"On the Structure of Space in General"*.

**Acknowledgements**

The authors would like to thank the Hebrew University of Jerusalem for permission to publish our translation of manuscript [1-115] and permission to display the excerpts shown in figures 1-4. Cormac O'Raifeartaigh thanks George Goulding of the WIT Reprographics Unit for assistance with the photographs of figure 3, and Professor John Stachel and Professor Jean Eisenstaedt for helpful discussions.

**Figure 1(a)**

Photograph of the first page of Einstein's manuscript *Über das sogenannte kosmologiche Problem,* document [1-115] on the Albert Einstein Online Archive (Einstein 1932a).

Hubbels Messungen an den extragalaktischen Nebeln haben nun ergeben, dass $\frac{1}{D}\frac{dD}{dt}$ $\left(=\frac{1}{A}\frac{dA}{dt}\right)$ für die Gegenwart eine Konstante $h$ ist. Sei $t$ die Gegenwart, so ist gemäss (6a)

$$t - t_0 = \frac{2}{3h} \quad \ldots (8)$$

Diese Zeit ergibt sich zu ungefähr $10^{10}$ Jahren. Natürlich wird zur Zeit die Dichte nicht wirklich unendlich gross gewesen sein; Land hat mit Recht darauf hingewiesen, dass für diese Zeit nur unsere rohe Approximation versagt, gemäss welcher die Dichte $\varrho$ als vom Orte unabhängig gesetzt ist.

Die Anwendung von (7) auf die Gegenwart liefert

$$3h^2 = \kappa \varrho c^2 (= 8\pi K \varrho) \quad \ldots (9)$$

Dies ist eine Relation zwischen der aus dem Doppler-Effekt ermittelten Hubbel-Konstante $h$ und der mittleren Dichte $\varrho$. Numerisch ergibt diese Gleichung für $\varrho$ die Grössenordnung $10^{-28}$, was mit den Schätzungen der Astronomen wohl vereinbar ist.

Aus diesen Betrachtungen geht hervor, dass wir beim heutigen Stande unserer Kenntnis die Thatsache einer von 0 verschiedenen Dichte der Materie nicht mit einer räumlichen Krümmung sondern mit einer Raum-Expansion in Zusammenhang zu bringen haben. Natürlich ist damit nicht gesagt, dass eine solche (positive oder negative) Krümmung nicht existiere. Für ihre Existenz liegen aber gegenwärtig keine Anhaltspunkte vor. Jedenfalls dürfte sie wesentlich kleiner sein, als die ursprüngliche Theorie (vgl. Gleichung 5) erwarten liess.

1-115

**Figure 1(b)**

Photograph of the last page of Einstein's manuscript *Über das sogenannte kosmologische Problem,* document [1-115] on the Albert Einstein Online Archive (Einstein 1932a).

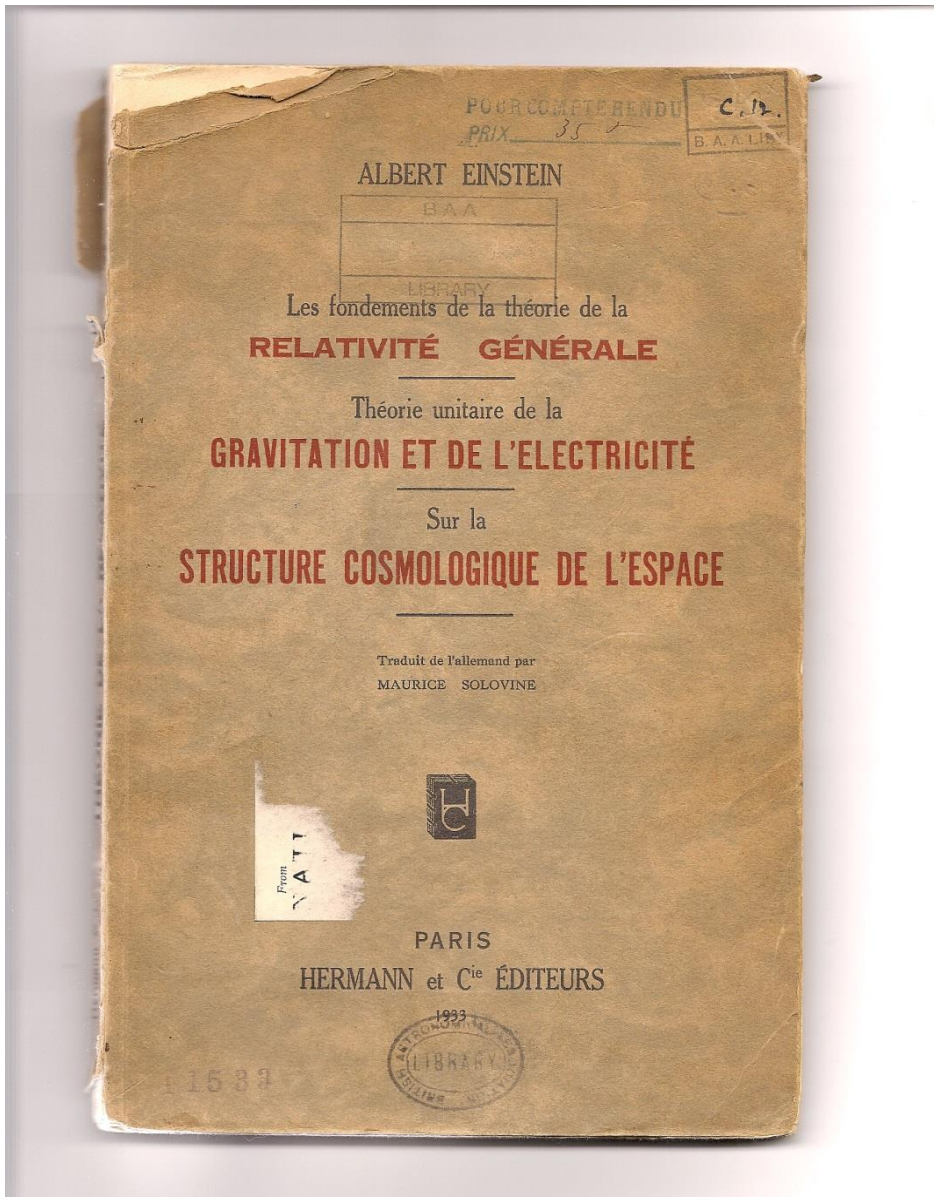

**Figure 2**

Photograph of a booklet of three Einstein papers translated into French by Maurice Solovine, published by Hermann et C$^{ie}$ in 1933. The collection includes the essay *Sur la Structure Cosmologique de l'Espace,* a translation of manuscript [1-115] (Einstein 1933).

**Figure 3(a)**

Photograph of the first page of the Solovine translation of Einstein's article (Einstein 1933).

(6 a) exprime, par conséquent, une expansion qui commence à un moment déterminé $t_0$. Il résulte de (7) que, pour cet instant, la densité est infinie. Les mesures effectuées par Hubble sur les nébuleuses extra-galactiques ont montré que

$$\frac{1}{D}\frac{dD}{dt} \left(= \frac{1}{A}\frac{dA}{dt}\right)$$

est pour le présent une constante $h$. Soit $t$ le présent, on a alors, conformément à (6 a),

(8) $$t - t_0 = \frac{2}{3h}.$$

Ce temps s'élève environ à $10^{10}$ années. Naturellement, à ce moment, la densité ne devait pas être réellement infiniment grande. Laue a fait remarquer avec raison que, pour ce temps, seule notre approximation grossière est en défaut, conformément à laquelle la densité $\rho$ est posée comme indépendante du lieu.

L'application de (7) au présent fournit

(9) $$3h^2 = \varkappa\rho c^2 \ (= 8\pi K\rho).$$

Ceci est une relation entre la constante $h$ de Hubble, déduite de l'effet Doppler, et la densité moyenne $\rho$. Numériquement, cette équation fournit pour $\rho$ une valeur de l'ordre $10^{-28}$, ce qui s'accorde bien avec l'estimation des astronomes.

Il résulte de ces considérations que, dans l'état actuel de nos connaissances, le fait d'une densité de la matière différente de zéro ne doit pas être relié théoriquement avec une courbure spatiale, mais avec une expansion spatiale. Nous ne voulons pas, naturellement, dire par là qu'une telle courbure (positive ou négative) n'existe pas. Mais nous n'avons, pour le moment, aucun indice de son existence. Elle devrait, en tout cas, être sensiblement moindre que ne faisait prévoir la Théorie primitive (cf. équation 5).

**Figure 3(b)**

Photograph of the last page of the Solovine translation of Einstein's article (Einstein 1933).

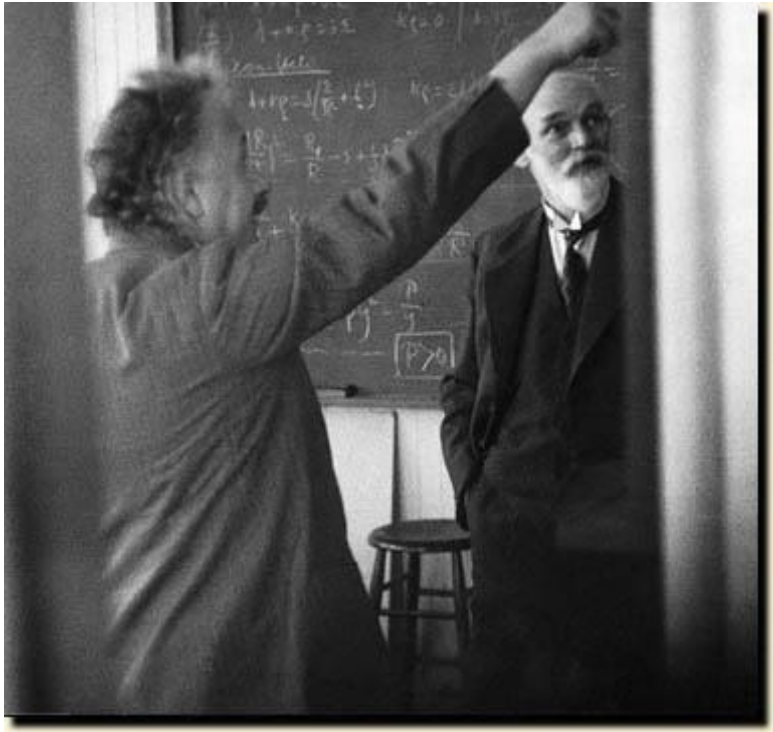

**Figure 4**

Einstein and de Sitter at work together at Caltech, Pasadena in 1932

**Appendix**

**On the so-called cosmological problem***

**(On the Cosmological Structure of Space)**

A. Einstein

When we call space and time in pre-relativistic physics "absolute", it has the following meaning. In the first instance, space and time, or the frame of reference, signify a reality in the same sense as, say, mass. Co-ordinates defined with respect to the chosen reference frame are immediately understood as results of measurement.[1] Propositions of geometry and kinematics are therefore understood as relations between measurements that have the significance of physical assertions which can be true or false. The inertial reference frame is understood to be a reality because its choice is inherent in the law of inertia.

Secondly, in terms of the laws obeyed, the physical reality denoted by the words space and time is independent of the behaviour of the rest of the physically real world, that is, independent of material bodies for example. According to classical theory, all relationships between measurements, which can themselves only be obtained using rulers and clocks, are independent of the distribution and motion of matter; the same is true for the inertial reference frame. Space enables the physical, in a sense, but cannot be influenced by the physical.

On account of the above state of affairs, some supporters of the theory of relativity have wrongly declared classical mechanics to be logically untenable. Such a theory is by no means logically untenable, although it is unsatisfactory from an epistemological point of view. In it, space and time play the role of an *a priori* reality, as it were, different from the reality of material bodies (and fields) which appear to some extent as a secondary reality. It is precisely this unsatisfactory division of physical reality that the general theory of relativity avoids. From a systematic point of view, the avoidance of this division of physical reality into two types is the main achievement of the general theory of relativity. The latter also made it possible to comprehend gravity and inertia from a common perspective. In view of the above, the use of general Gaussian co-ordinates, which provide a continuous labelling of points in space-time without reference to metric relations, is a mere (albeit indispensable) tool that allows the metric properties of the continuum to be coordinated with its other properties (gravitational field, electromagnetic field, law of motion).[2]

---

* Translation from the original manuscript *'Über das sogenannte kosmologische Problem'*, document [1-115] on the Albert Einstein Online Archive (transl. C. O'Raifeartaigh, B. McCann and W. Nahm). The article was published in French in 1933 under the title *'Sur la Structure Cosmologique de l'Espace'* in the book *'Les Fondaments de la Théorie de la Relativité Générale'*, Hermann et C$^{ie}$ (transl. M. Solovine).

[1] At least this will be true if one considers ideal rulers and clocks realisable in principle.

Since, according to the general theory of relativity, the metric properties of space are not given in themselves but are instead determined by material objects that force a non-Euclidean character on the continuum, a problem arises that is absent from the classical theory. Namely, since we may assume that the stars are distributed with a finite density everywhere in the world, that is, a non-zero average density of matter in general, there arises the question of the influence of this mean density on the (metric) structure of space on a large scale; this is the so-called cosmological problem that we wish to address in this short note. For simplicity, we will ignore the fact that matter is concentrated in stars and star systems, separated by apparently empty regions, but instead treat matter as though it were continuously distributed over astronomically large regions.

Moreover, the assumption of a finite non-zero mean density of matter already leads to difficulties from the point of view of the Newtonian theory, as the astronomers have long known. Namely, according to the theorem of Gauss, the number of lines of gravitational force that cross a closed surface from the outside to the inside is equal to a constant multiple of the gravitating mass enclosed by the surface. If this matter has the constant density $\rho$, then for a sphere of radius $P$, the number of lines of force is proportional to $P^3$. Therefore the flux of force per unit area of the sphere is proportional to the radius of the sphere, and so the greater the radius of the sphere, the greater it will be. Hence, according to Newton's theory, free matter of a finite constant density cannot remain in global equilibrium. To avoid the resulting difficulty, the astronomer Seeliger proposed a modification of the Newtonian law of attraction for large distances. Of course, this question had nothing to do with the problem of space.

The corresponding problem in the general theory of relativity leads to the question: how is it possible to have a space with a spatially constant density of matter that is at rest relative to it? Such a space should be considered a crude idealization for a theoretical treatment of the actual space-time-continuum.

---

[2] This is clearly seen when one treats the classical theory using general coordinates. The Riemann tensor of rank four $R_{ik,lm}$ is then set equal to zero. This implies a complete determination of the metric field of the $g_{\mu\nu}$, a determination that does not involve the other physical variables. From the point of view of the general covariant description, this is how the absolute character of time and space is expressed in classical theory.

According to the general theory of relativity, the metric or gravitational field described by the $g_{\mu\nu}$ is related to the energy or mass density tensor $T_{\mu\nu}$ by the equations

$$R_{\mu\nu} - \frac{1}{2} g_{\mu\nu} R = -\kappa\, T_{\mu\nu} \quad \ldots \quad (1)$$

Here $R_{\mu\nu}$ signifies the once-contracted Riemann tensor

$$\left. \begin{array}{l} R_{\mu\nu} = -\Gamma^{\alpha}_{\mu\nu,\alpha} + \Gamma^{\alpha}_{\mu\alpha,\nu} + \Gamma^{\alpha}_{\mu\beta}\Gamma^{\beta}_{\nu\alpha} - \Gamma^{\alpha}_{\mu\nu}\Gamma^{\beta}_{\alpha\beta} \\ \Gamma^{\alpha}_{\mu\nu} = g^{\alpha\mu}\{^{\alpha}_{\mu\nu}\} = g^{\alpha\beta}\{^{\mu\nu}_{\beta}\} = \frac{1}{2} g^{\alpha\beta}\left(g_{\mu\alpha,\nu} + g_{\nu\alpha,\mu} - g_{\mu\nu,\alpha}\right), \end{array} \right\} \quad (1a)$$

where the usual differentiation is denoted by a comma.

If "matter" can be idealised as pressure-free and the influence of effects other than gravity can be neglected, then

$$T^{\mu\nu} = \rho u^{\mu} u^{\nu} \quad \ldots \quad (2)$$

where $u^{\mu}$ denotes the contravariant velocity four-vector $\frac{dX_{\mu}}{d\tau}$ and $\rho$ the scalar density of matter.[3] Naturally, it is also assumed that the energy density of the ponderable matter outweighs that of the radiation to such an extent that the latter can be neglected. Although the validity of this assumption is not entirely assured, the approximation introduced does not essentially alter the results.

One sees first of all that a world with a non-zero mean density of matter cannot be Euclidean. For such a world is given in terms of the special theory of relativity by a line element

$$g_{\mu\nu} dx^{\mu} dx^{\nu} = ds^2 = dx_1^2 + dx_2^2 + dx_3^2 - c^2 dt^2 \quad (3)$$

i.e., by constant values of the $g_{\mu\nu}$. $R_{\mu\nu}$ and $R$ then vanish and with them the left-hand side of (1). It follows that the right-hand side of (1) must also vanish, and with it $\rho$, in contradiction to our assumption.

After Euclidean space, the simplest spatial structure conceivable would seem to be one that is static (all $g_{\mu\nu}$ independent of $t$) and that has constant curvature with respect to the "spatial" sections ($t$ = constant).

As is well-known, a three-dimensional space with constant positive curvature (in particular a "spherical" space) is characterised by a line element $d\sigma^2$ of the form

---

[3] $d\tau$ is the element of eigentime, thus when the spatial $ds^2$ is positive $d\tau^2 = -ds^2$.

$$d\sigma^2 = \frac{dx_1^2 + dx_2^2 + dx_3^2}{\left(1 + \frac{r^2}{(2P)^2}\right)^2} \qquad (r^2 = x_1^2 + x_2^2 + x_3^2)$$

where the point $x_1 = x_2 = x_3 = 0$ is only apparently singled out.[4]

*[A world that is static and spatially spherical is therefore described by the line element

$$ds^2 = \frac{dx_1^2 + dx_2^2 + dx_3^2}{\left(1 + \frac{r^2}{(2P)^2}\right)^2} - c^2 dt^2 \qquad (3a)$$

It seems plausible *a priori* (as regards properties of symmetry) that in such a world, one could have matter at rest ($u^1 = u^2 = u^3 = 0$) and of constant density over space and time. In addition one has

$$u^4 = \frac{dt}{d\tau} = \frac{1}{c} \; ;$$

---

[4] This is most easily seen by embedding a three-dimensional sphere in a 4-dimensional Euclidean space with Cartesian coordinates $\xi_1, \xi_2, \xi_3, \xi_4$ and centre $0,0,0,-P$:

$$\xi_1^2 + \xi_2^2 + \xi_3^2 + (\xi_4 + P)^2 = P^2$$

is then the equation of the sphere and the line element on it is measured as

$$d\xi_1^2 + d\xi_2^2 + d\xi_3^2 + d\xi_4^2 = d\sigma^2.$$

Using the equation of the sphere, one of the four coordinates and coordinate differentials can be eliminated. (Introduction of three coordinates on the sphere instead of the four coordinates $\xi_1, \ldots, \xi_4$). This is best realised (avoidance of square roots) by the "stereographic projection" of the points on the sphere onto the hyperplane $\xi_4 = -2P$, in accordance with the accompanying sketch.

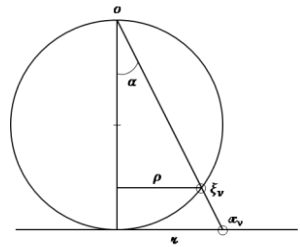

$$\rho^2 = \xi_1^2 + \xi_2^2 + \xi_3^2 \; : \; r^2 = x_1^2 + x_2^2 + x_3^2$$

The $\xi_\nu$ are replaced by $x_\nu$ in accordance with the relation:

$$\frac{x_\nu}{\xi_\nu} = \frac{r}{\rho} = \frac{1}{\cos^2\alpha} = 1 + \tan^2\alpha = 1 + \frac{r^2}{(2P)^2} \quad (\nu = 1..4).$$

This yields $\xi_1, \xi_2, \xi_3, \xi_4$ ($\xi_4 = -2P$) as functions of the $x_\nu, \nu = 1,\ldots,3$, from which one finds by differentiation the $d\xi$, and hence $d\sigma^2$ as a function of the $x_\nu$ and the $dx_\nu$, in accordance with the formula given in the text.

and hence: $u_4 = g_{44}u^4 = c$.

In consequence, one must insert 0 for the $T_{\mu\nu}$ in the second term of (1), up to $T_{44}$. Only $T_{44} = \rho u_\mu u_\nu = \rho c^2$ is different from 0.

From 3(a), in accordance with (1a) one calculates for the $R_{\mu\nu}$ ($x_1 = x_2 = x_3 = 0$) the values

$$\begin{pmatrix} -\frac{2}{P^2} & 0 & 0 & 0 \\ 0 & -\frac{2}{P^2} & 0 & 0 \\ 0 & 0 & -\frac{2}{P^2} & 0 \\ 0 & 0 & 0 & 0 \end{pmatrix},$$

for $R_{\mu\nu} - \frac{1}{2}g_{\mu\nu}R$, the values

$$\begin{pmatrix} \frac{1}{P^2} & 0 & 0 & 0 \\ 0 & \frac{1}{P^2} & 0 & 0 \\ 0 & 0 & \frac{1}{P^2} & 0 \\ 0 & 0 & 0 & -\frac{3c^2}{P^2} \end{pmatrix};$$

while for $-\kappa T$ one obtains the values

$$\begin{pmatrix} 0 & 0 & 0 & 0 \\ 0 & 0 & 0 & 0 \\ 0 & 0 & 0 & 0 \\ 0 & 0 & 0 & -\kappa\rho c^2 \end{pmatrix}]^*$$

Thus from (1) the two contradictory equations are obtained

$$\left.\begin{matrix} \frac{1}{P^2} = 0 \\ \frac{3c^2}{P^2} = \kappa\rho c^2 \end{matrix}\right\} \quad (4)$$

Therefore equations (1) do not allow the possibility of a non-zero uniform density of matter $\rho$. This immediately creates a serious difficulty for the general theory of relativity, given that time-independent spatial structures other than those given by (3a) (for positive or negative $P^2$) are inconceivable.

I initially found the following way out of this difficulty. The requirements of relativity permit and suggest the addition of a term of the form $\lambda g_{\mu\nu}$ to the left hand side of (1), where $\lambda$ denotes a universal constant (cosmological constant), which must be small enough that the additional term need not be considered in practice when calculating the sun's gravitational field and the motion of the planets. Completed in this manner, the equations are

$$\left( R_{\mu\nu} - \frac{1}{2} g_{\mu\nu} R \right) - \lambda g_{\mu\nu} = -\kappa\, T_{\mu\nu} \quad \ldots (1b)$$

Instead of equations (4), one then finds

$$\left. \begin{array}{l} \frac{1}{P^2} = \lambda \\ \frac{3c^2}{P^2} = -\lambda c^2 + \kappa\rho c^2 \end{array} \right\} (4a)$$

These equations are consistent and yield the following value for the world radius

$$P = \frac{2}{\sqrt{\kappa\rho}} = \frac{c}{\sqrt{2\pi K\rho}} \quad (5)$$

where $K$ denotes the gravitational constant as measured in the usual system of measurement.

However, it later emerged as a result of research by Lemaitre and Friedmann that this resolution of the difficulty is unsatisfactory for the following reason.

The afore-mentioned authors also proceeded from equations (1b). However, they generalized the approach (3a) by introducing the world radius $P$ (and the density $\rho$) not as a constant, but as an *a priori* unknown function of time. Equations (1b) then show that the solution (4a), (5) has an unstable character. This means that, for solutions that differ only slightly from (4a) at a particular point in time, $P$ does not oscillate about the value given by (5) but instead deviates (for larger or smaller values of time) more and more from the value of $P$ given by (5). Furthermore, if one adopts these "dynamic" solutions to the problem, then both the magnitude and the sign of $\lambda$ will remain undetermined, and indeed even the sign of $\frac{1}{P^2}$, so that negative spatial curvatures also appear possible[5] and thus the basis for the postulate of a spatially closed world is completely removed.

Given that the theory leads us to adopt dynamic solutions for the structure of space, it is no longer necessary to introduce the universal constant $\lambda$, as there are dynamic solutions for (1) of type (3a) for which $\lambda = 0$.

---

[5] Heckmann was the first to point this out.

In recent times, the resolution of the problem has experienced a strong stimulus from empirical results in astronomy. Measurements of the Doppler effect (in particular those of Hubbel) of the extra-galactic nebulae, which have been recognized as similar formations to the Milky Way, have shown that the further these formations are from us, the greater the velocity with which they hasten away. Hubbel's investigations also showed that these formations are distributed in space in a statistically uniform manner, giving empirical support to the underlying theoretical assumption of a uniform mean density of matter. The discovery of the expansion of the extra-galactic nebulae justifies the shift to dynamic solutions for the structure of space, a step that heretofore would have appeared to be an expedient justified only by theoretical necessity.

Thus the theory can now, without the introduction of a $\lambda$-term, accommodate a finite (mean) density of matter $\rho$ on the basis of equations (1), using relation 3(a) with $P$ (and $\rho$) variable over time. Here, it should be noted that it is not the coordinates $x_1, x_2, x_3$ of a particle that remain constant over time, but instead the quantities $\frac{x_1}{P}, \frac{x_2}{P}, \frac{x_3}{P}$, as is seen from a straightforward geometric argument. We will not introduce these quantities themselves as new coordinates, but instead the quantities $P_0 \frac{x_1}{P}, P_0 \frac{x_2}{P}, P_0 \frac{x_3}{P}$, where $P_0$ signifies a length of the order of magnitude of the "world radius". We do this to ensure that differences between coordinates will be of the same order of magnitude as lengths measured with a ruler.[6] If we once more label these new co-ordinates as $x_1, x_2, x_3$ and in this new system again let $r = \sqrt{x_1^2 + x_2^2 + x_3^2}$, then the relation (3a) takes the form

$$ds^2 = \left(\frac{P}{P_0}\right)^2 \frac{dx_1^2 + dx_2^2 + dx_3^2}{\left(1 + \frac{r^2}{(2P_0)^2}\right)^2} - c^2 dt^2 \quad \ldots (3b)$$

We can regard $P_0$ as the world radius $P$ at a particular point in time $t_0$. Only the "expansion factor" $\frac{P}{P_0}$ ($= A$) is then variable over time.

We have already noted that, if we take $A$ to be constant over time, i.e., without an "expansion" of space, we cannot explain a constant density of matter $\rho$ solely by the assumption of a curvature of space. On the other hand, it will be shown that the existence of a finite density $\rho$ does not in any way demand the existence of a (three-dimensional) curvature of space. This amounts to replacing relation (3b) by

$$ds^2 = A^2(dx_1^2 + dx_2^2 + dx_3^2) - c^2 dt^2 \quad \ldots \quad (3c)$$

where $A$ is a function of $t$ (= $x_4$) alone. Introducing this relation in (1) gives

$$2A \frac{d^2 A}{dt^2} + \left(\frac{dA}{dt}\right)^2 = 0 \quad \ldots \quad (6)$$

$$3 \left(\frac{\frac{dA}{dt}}{A}\right)^2 = \kappa \rho c^2 \quad \quad (7)$$

---

[6] A very thorough examination of the general problem and its various special cases based on this choice of coordinates has been carried out by Tolman. In addition, De Sitter has given a clear and exhaustive account of all possible cases.

Equation (6) yields

$$A = c(t - t_0)^{2/3} \quad \ldots \quad (6a)$$

If $l$ is the time-independent distance $\sqrt{\Delta x_1^2 + \Delta x_2^2 + \Delta x_3^2}$ between two masses measured in terms of coordinates, then according to (3c), $Al$ is the distance $D$ between these two mass points as measured with a ruler. (6a) thus expresses an expansion that begins at a particular time $t_0$. For this point in time (7) shows that the density is infinite. Hubbel's measurements of the extra-galactic nebulae have shown that for the present, $\frac{1}{D}\frac{dD}{dt}$ ($= \frac{1}{A}\frac{dA}{dt}$) is a constant $h$. If $t$ is the present time, then according to (6a)

$$t - t_0 = \frac{2}{3h} \quad \ldots (8)$$

This time-span works out at approximately $10^{10}$ years. Of course, at that time the density will not actually have been infinitely large; Laue has rightly pointed out that our rough approximation, according to which the density $\rho$ is independent of location, breaks down for this time.

Applying (7) to the present yields

$$3h^2 = \kappa \rho c^2 \ (= 8\pi K \rho) \ldots (9)$$

This is a relation between the Hubbel constant $h$, determined from the Doppler effect, and the mean density $\rho$. Numerically, this equation gives an order of magnitude of $10^{-28}$ for $\rho$, which is not incompatible with the estimates of the astronomers.

It follows from these considerations that in the light of our present knowledge, the fact of a non-zero density of matter need not be reconciled with a curvature of space, but instead with an expansion of space. Of course, this does not mean that such a curvature (positive or negative) does not exist. However, there is at present no indication of its existence. In any case, it may well be substantially smaller than might have been suggested by the original theory (see equation 5).

**Translation notes**

(i) We have preserved the layout of manuscript [1-115] in terms of paragraph structure, numbering for equations and footnotes.
(ii) On the fifth page of manuscript [1-115], the unnumbered equation immediately following equation (3) and associated footnote (4) is written in Walther Mayer's handwriting.
(iii) On the fifth and sixth pages, the passage enclosed in square brackets and marked with an asterisk is missing from manuscript [1-115]. We obtained this passage from the Solovine translation.
(iv) The name Hubble is misspelt as Hubbel each time it occurs in the manuscript, while Lemaître is written as Lemaitre. We have retained these misspellings for authenticity.
(v) In equations (5) and (9), it is clear from the algebra that the Einstein constant $\kappa$ is taken as $8\pi G/c^2$ while the parameter $K$ is the gravitational constant $G$.
(vi) Einstein does not state what value he assumes for the Hubble constant $h$ in the calculations arising from equations (8) and (9). Comparison with the Einstein-de Sitter paper of 1932 suggests that he used a value of $h = 500$ km s$^{-1}$ Mpc$^{-1}$, the standard value at the time.
(vii) Units of measurement are not given for the density estimate $\rho = 10^{-28}$. A comparison with the Einstein-de Sitter paper of 1932 suggests that the units are g cm$^{-3}$.
(viii) Equation (8) implies a value of 1.3 billion years rather than "approximately 10 billion years" as stated by Einstein.